\documentclass[prb,twocolumn,showpacs,amsmath,amssymb,superscriptaddress,longbibliography]{revtex4-1}
\usepackage{graphicx}
\usepackage{breqn}
\usepackage{xcolor}
\usepackage{color}
\usepackage{booktabs}
\usepackage{float}
\usepackage{longtable}
\usepackage{epsfig}
\usepackage{bm}
\usepackage{dcolumn}
\usepackage{lipsum, babel}
\usepackage{soul}

\usepackage{rotating} 

\usepackage[normalem]{ulem}
\usepackage{epstopdf}
\usepackage{natbib}
\usepackage[bookmarksopen, colorlinks,linkcolor=blue,citecolor=blue,urlcolor=blue]{hyperref}

\makeatletter
\let\cat@comma@active\@empty
\makeatother
\begin{document}

\title{Altermagnets and beyond: Nodal magnetically-ordered phases}

\author{Tomas~Jungwirth}
\email{jungw@fzu.cz}
\affiliation{Institute of Physics, Czech Academy of Sciences, Cukrovarnick\'a 10, 162 00, Praha 6, Czech Republic}
\affiliation{School of Physics and Astronomy, University of Nottingham, NG7 2RD, Nottingham, United Kingdom}

\author{Rafael M. Fernandes}
\affiliation{Department of Physics, University of Illinois Urbana-Champaign, Urbana, IL 61801, USA}

\author{Jairo Sinova}
\affiliation{Institut f\"ur Physik, Johannes Gutenberg Universit\"at Mainz, D-55099 Mainz, Germany}

\author{Libor \v{S}mejkal}
\email{lsmejkal@pks.mpg.de}
\affiliation{Max Planck Institute for the Physics of Complex Systems, N\"othnitzer Str. 38, 01187 Dresden, Germany}
\affiliation{Institute of Physics, Czech Academy of Sciences, Cukrovarnick\'a 10, 162 00, Praha 6, Czech Republic}
\affiliation{Institut f\"ur Physik, Johannes Gutenberg Universit\"at Mainz, D-55099 Mainz, Germany}

\date{\today}

\begin{abstract}
The recent discovery of altermagnets has opened new perspectives in the field of ordered phases in condensed matter.  In strongly-correlated superfluids, the nodal p-wave and d-wave ordered phases of $^{3}$He and cuprates play a prominent role in physics for their rich phenomenology of the symmetry-breaking order parameters. While the p-wave and d-wave superfluids have been extensively studied over the past half a century, material realizations of their magnetic counterparts have remained elusive for many decades.  This is resolved in altermagnets, whose recent discovery was driven by research in the field of spintronics towards highly scalable information technologies. Altermagnets  feature d, g or i-wave magnetic ordering, with a characteristic alternation of spin polarization and spin-degenerate nodes. Here we review how altermagnetism can be identified from symmetries of collinear spin densities in crystal lattices, and can be realized  at normal conditions in a broad family of insulating and conducting materials.  We highlight salient electronic-structure signatures of the altermagnetic ordering, discuss extraordinary relativistic and topological phenomena that emerge in their band structures, and comment on strong-correlation effects. We then extend the discussion to non-collinear spin densities in crystals, including the prediction of p-wave magnets, and conclude  with a brief summary of the reviewed  physical properties  of the nodal magnetically-ordered phases. 
\end{abstract}

\maketitle

\subsection{Introduction}

\begin{figure}[h!]
	\centering
	\includegraphics[width=1\linewidth]{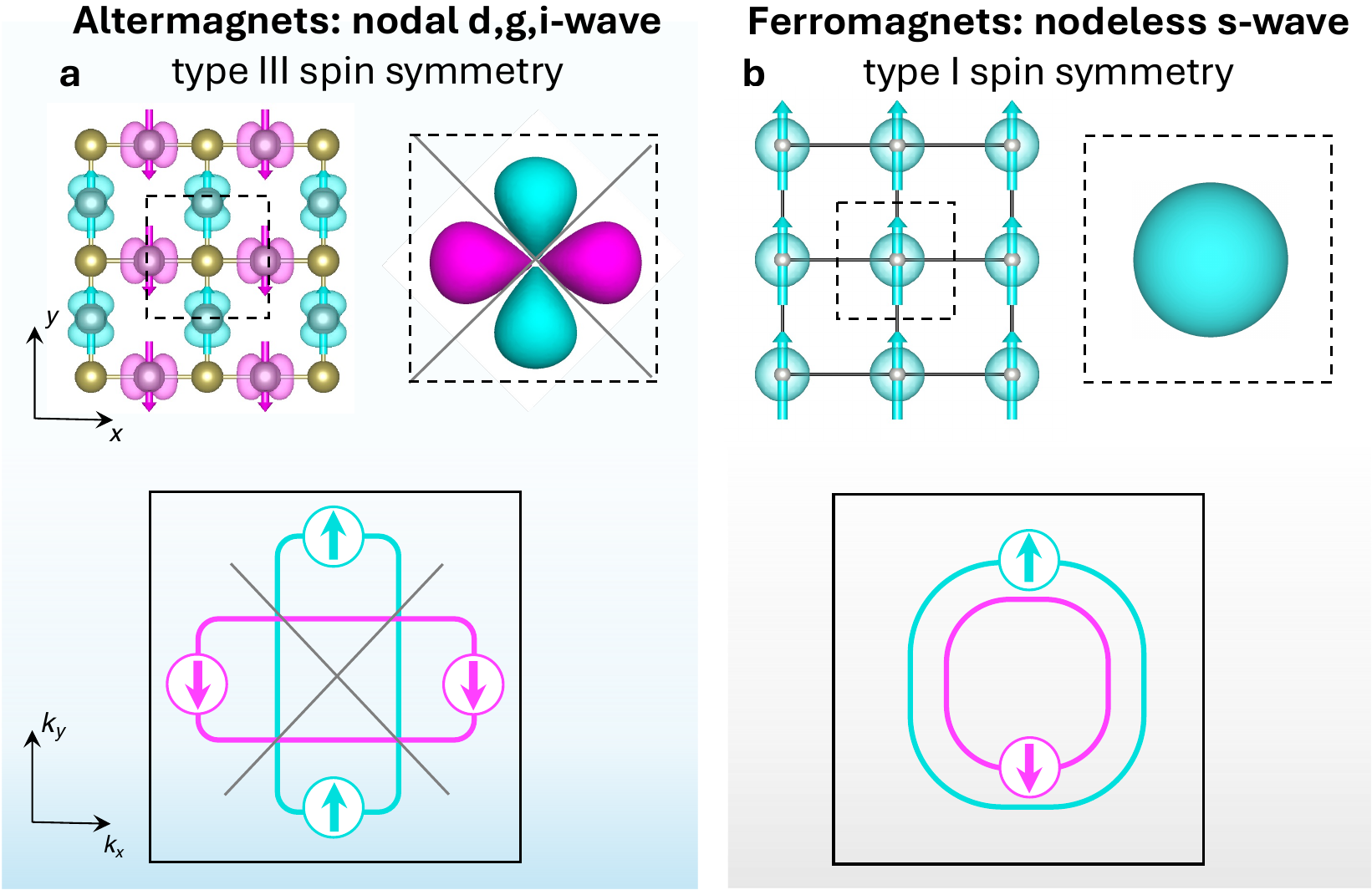}
	\caption{\textbf{Altermagnets and ferromagnets.}
		{\bf a,} Nodal altermagnetic ordering in the crystal space (top) and in the momentum-space energy iso-surfaces/Fermi surfaces (bottom). The cartoons illustrate a d-wave ordering. Spin-degenerate nodes are highlighted by black lines in the 2D cartoons. Arrows and corresponding colors represent opposite spin orientations. {\bf b,} Nodeless s-wave ferromagnetic ordering in crystal (top) and momentum (bottom) space. In superconductors (superfluids), the corresponding d-wave and s-wave phases are manifested in the quasiparticle gap function.
}
\label{fig:AM-FM}
\end{figure}

In research, advances in one field can bring new perspectives on open problems in a distant field.  Altermagnets\cite{Smejkal2021a} are an example of such a scenario. Their discovery, originally driven by spintronics research, opens new chapters in the distant field of broken-symmetry phases with nodal p-wave, d-wave or other beyond s-wave character of ordering. Specifically, recent developments in spintronics motivated a search for magnetically-ordered materials featuring a combination of strong spin-polarized transport phenomena, typical of ferromagnets, with high spatial, temporal and energy scalability enabled by the absence of net magnetization, characteristic of antiferromagnets. Altermagnets were predicted to combine these seemingly mutually exclusive favorable properties in one material. This is thanks to the even-parity-wave nodal character of their spin-polarized electronic structure in the momentum-space, and the corresponding  nodal spin-density in the crystal space. Such a magnetic phase can be generated by an ordering of mutually rotated neighboring spins (by 180$^\circ$), combined with mutually rotated atoms (nuclei with electronic orbitals) in the crystal lattice that carry the spins, as illustrated in  Fig.~\ref{fig:AM-FM}a. The seemingly contradictory nature of altermagets, combining strong spin polarization with a vanishing magnetization, arises from specific symmetries of collinear spin densities in crystals, rendering altermagnets an exclusively distinct symmetry class from the traditional collinear ferromagnets and antiferromagnets. 

Earlier review articles\cite{Smejkal2022a,Smejkal2022AHEReview} already elaborated in detail on the spintronic drivers of the altermagnetic research. They also covered the basic symmetry delineation of this third distinct class of collinear magnets, which can be cast on formal grounds using the concept of spin symmetry groups.

In this review, we focus on the nodal even-parity-wave nature of magnetic ordering in altermagnets, its signatures and material realizations, and implications for  extraordinary electronic-structure phenomena. To set the stage, we start in Sec.~\ref{phases} by recalling the established notions of interacting fermions forming conventional nodeless s-wave ordered phases in superfluids and ferromagnets (Fig.~\ref{fig:AM-FM}b). We highlight that these two archetype ordered phases in condensed-matter physics share a common theoretical foundation provided by the formalism of Fermi-liquid instabilities. We will then use the formalism for the introduction into physics of p-wave and d-wave broken-symmetry phases, which are well established in the superfluids. By this, on one hand, we recall the elusive nature of counterpart material realizations of p-wave and d-wave magnetic Fermi-liquid instabilities. On the other hand, it enables us to contrast the Fermi-liquid formalism with the spin-symmetry approach, where the latter has led to the identification of a broad family of material candidates of d, g, and i-wave altermagnetic phases. The key difference is that the Fermi liquid formalism focuses on the Fermi surface instabilities and corresponding ordered phases in the momentum space. In contrast, the spin-symmetry approach is based on  symmetries of  spin densities in the crystal-structure direct space, which are then reflected in the corresponding symmetries in the electronic-structure momentum space. This approach enables to capture the interplay of single-particle potentials of the crystal lattice and electron-electron interactions, which is essential for the formation of  the altermagnetic phases in the broad family of identified materials. 

The field of altermagnetism is rapidly gaining momentum with more than three hundred studies reported over the past two years. In Sec.~\ref{realizations} we give an overview of salient band-structure characteristics and summarize identified physical principles enabling material realizations of the altermagnetic ordering. In Sec.~\ref{relativistic} we include a discussion of relativistic spin-orbit-coupling  and topological phenomena in the electronic structure, and comment on effects of strong correlations. In Sec.~\ref{odd-parity} we extend the review to non-collinear spin densities in crystal lattices, including the prediction of materials with p-wave magnetic ordering.  We briefly summarize the review in Sec.~\ref{summary}.

\subsection{Symmetry-breaking ordered phases} 
\label{phases}
Throughout the century of quantum condensed-matter physics, interacting fermions have provided a fascinating research playground despite, or perhaps because they represent a generally unsolvable many-body problem. Breakthroughs in the field thus often relied on applying  approaches that were unorthodox at the time. Among those, the Landau's Fermi-liquid theory of metals \cite{Landau1957,Vignale2022} is arguably the first  that comes to mind. It sidestepped the problem of the interacting many-body ground state by focusing, instead, on two-body interactions involving elementary quasiparticle excitations near the Fermi surface, whose lifetime becomes infinite when approaching the Fermi surface. 
In certain systems, instabilities of the Fermi surface induced by  interactions between these quasiparticles can trigger transitions from normal to ordered symmetry-breaking phases. 

Conventional BCS superconductivity and itinerant ferromagnetism are the primary examples  in which interactions lead to an ordered state rendering the Fermi surface unstable. In the former case, below the transition temperature, an attractive Cooper-pairing interaction (mediated by phonons), even if infinitesimally weak, causes the formation of a coherent macroscopic quantum state of bound pairs of fermion quasiparticles. This is manifested as a quasiparticle excitation gap in place of the normal-state Fermi surface\cite{Bardeen1957}. In the case of ferromagnetism, a Pomeranchuk instability \cite{Pomeranchuk1959} is driven by an exchange interaction resulting from the combination of the electron-electron Coulomb repulsion and Pauli's exclusion principle\cite{Heisenberg1926,Dirac1926}. When this interaction exceeds a critical value \cite{Pomeranchuk1959},  the normal-state spin-degenerate Fermi surface splits into majority-spin (spin-up) and minority-spin (spin-down) Fermi surfaces. These superconducting and magnetic phases break the U(1) gauge (particle-conservation) symmetry and the SO(3) spin-rotation symmetry, respectively, while preserving the crystal-lattice symmetry in the momentum-dependent energy spectra. In other words, the quasiparticle excitation gap of the former and the splitting between the spin-up and spin-down Fermi surfaces of the latter are principally isotropic in momentum space (more precisely, respect the rotation symmetries of the underlying crystal lattice). Accordingly, they are referred to as nodeless s-wave (angular momentum $l=0$) ordered phases \cite{Sigrist1991,Annett1995,Houzet2012,Tsuei2000,Mackenzie2003,Monthoux2007}.

\begin{figure}[h!]
	\centering
	\includegraphics[width=1\linewidth]{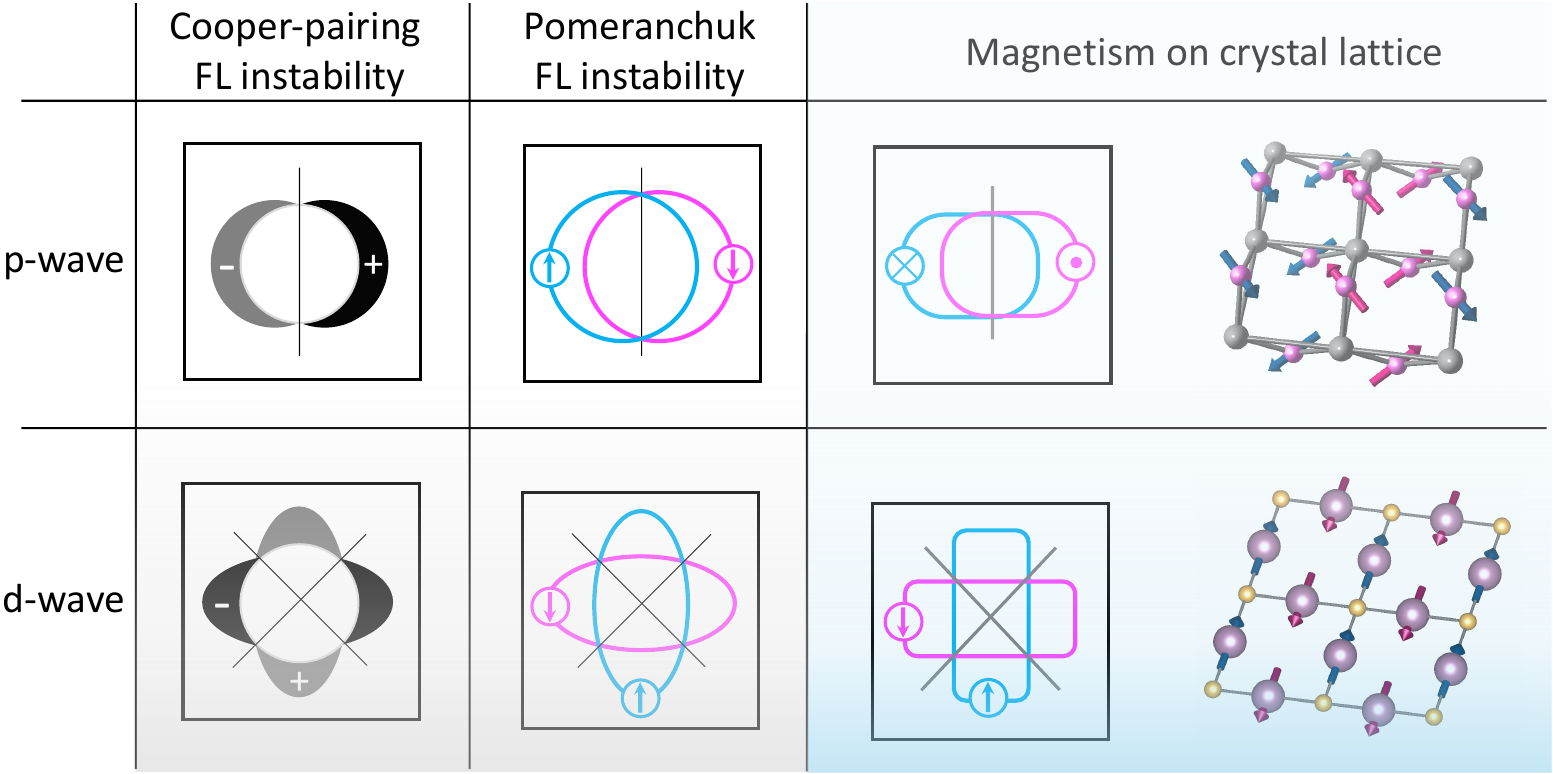}
	\caption{
	\textbf{Nodal ordering in superfluids and magnets.}
Schematic comparison of the Cooper-pairing Fermi-liquid instabilities with p-wave and d-wave gap functions ($\pm$ refers to the phase of the order parameter), their searched-for counterpart magnetic Pomeranchuk Fermi-liquid instabilities, and the p-wave and d-wave magnetic ordering identified from symmetries of spin densities in crystal lattices. 		
}
\label{fig:SC-M}
\end{figure}

Cooper-paring  instabilities, leading to ordered phases characaterized by additional symmetry breaking beyond the broken gauge symmetry, have revealed a rich phenomenology of their  order parameters \cite{Sigrist1991,Annett1995,Houzet2012,Tsuei2000,Mackenzie2003,Monthoux2007}.  Prominent examples are the superfluidity of $^{3}$He with an odd-parity p-wave ($l=1$) gap function (Fig.~\ref{fig:SC-M}) \cite{Leggett1975}, and the unconventional superconductivity of cuprates with an anisotropic d-wave ($l=2$) gap function (Fig.~\ref{fig:SC-M}).

Besides s-wave ferromagnetism, Pomeranchuk instabilities could also generate ordered phases with $l>0$ odd or even-parity distortions of the Fermi surface. Both spin-polarized and spin-unpolarized $l>0$ instabilities are possible. These  $l>0$  Pomeranchuk instabilities  have been more elusive, despite the long history of their research. Already in 1940's, a phase corresponding to a spin-unpolarized p-wave ($l=1$) Pomeranchuk instability, resulting in a spontaneous equilibrium charge current from a shifted parity-breaking Fermi surface, was among the unsuccessful early attempts to explain superconductivity \cite{Born1948}, before the successful BCS theory\cite{Bardeen1957}. Over the following eight decades, theoretical studies of both spin-unpolarized and spin-polarized (Fig.~2) p-wave Pomeranchuk instabilities   in correlated Fermi liquids have advanced the theoretical understanding of these phases,  including the constraints on the possible formation of the corresponding ordered many-body ground states with spontaneous charge or spin currents \cite{Born1948,Bohm1949,Hirsch1990,Wu2004,Wu2007,Jung2015,Kiselev2017,Wu2018,Klein2019}. 

Unlike the case of $l=1$ instabilities, a nematic phase, which is characterized by the spontaneous breaking of rotational symmetry and is linked to a spin-unpolarized $l=2$ even-parity-wave Pomeranchuk instability, has been observed in various correlated systems and proposed to emerge via a variety of  microscopic mechanisms\cite{Kivelson1998,Lilly1999,Oganesyan2001,Borzi2007,Fradkin2007,Fradkin2010,Fernandes2014,Lee2018b,Quintanilla2023}.
Spin-polarized $l=2$ Pomeranchuk instabilities (Fig.~\ref{fig:SC-M}) have been also theoretically discussed  \cite{Kivelson2003,Wu2007,Wu2018}, however, their physical realizations have remained elusive similarly to the case of the odd-parity-wave Pomeranchuk instabilities.  The difficulty is that the Landau Fermi liquid parameter, characterizing the strength of the Landau  interaction function in the given $l>0$ angular-momentum component,  has to overcome the threshold for the instability. Simultaneously, the other $l$-component Fermi liquid parameters should remain below the threshold for their respective instabilities (namely the $l=0$ ferromagnetic instability), to avoid a formation of the competing phase with a different symmetry breaking.

The recent discovery of altermagnets with the $l>0$ even-parity-wave magnetic order was based on a principally distinct unorthodox approach sidestepping the unknown exact description of the interacting many-body problem.  Instead of focusing on Fermi-surface  instabilities of correlated metallic Fermi liquids, the anisotropic symmetry breaking of the spin-polarized energy iso-surfaces of altermagnets was found to be promoted  by suitable  non-relativistic symmetries of spin densities in crystal-lattices (Fig.~\ref{fig:SC-M}). These result from the interplay of single-particle potentials of the crystal lattice, which can originate from  mutually rotated neighboring atoms,  and electron-electron interactions, which lead to the antiparallel alignment of the  neighboring spins  \cite{Smejkal2021a,Smejkal2022a}. 

Here we point out that unlike the commonly employed relativistic magnetic symmetries of coupled real and spin space, the non-relativistic spin symmetries consider pairs of generally different  transformations in the real space and the spin space. The spin symmetries can thus classify and describe a much richer landscape of magnetically ordered phases. Consequently, altermagnetism was theoretically identified in metals and insulators, often at ambient conditions \cite{Smejkal2022a}. Experimental observations of altermagnetic band structures, predicted by the  spin-symmetry analysis  and density-functional-theory (DFT) calculations, have been recently reported in photoemission experiments  \cite{Krempasky2024,Lee2024,Osumi2024,Hajlaoui2024,Reimers2024}. 

Further analysis of the spin symmetries in crystal-lattices, supported by DFT calculations, has  also led to the prediction of material candidates hosting the  odd-parity-wave magnetic phases \cite{Hellenes2023} (Fig.~\ref{fig:SC-M}). As in the case of altermagnets, the odd-parity wave magnetism realized in suitable lattice and spin structures does not require strong correlations and extreme external conditions.   


\subsection{Salient signatures and physical realizations of nodal magnetic ordering in altermagnets}
\label{realizations} 
In this section, we focus on the altermagnetic ordering in the ground state which spontaneously breaks symmetries of the Hamiltonian describing the interacting electron system in an ionic crystal lattice. We omit the perturbatively weak symmetry-breaking effects of the relativistic spin-orbit-coupling single-particle term in the Hamiltonian. These will be discussed in the following section. 

Fig.~\ref{fig:AM-FM}b illustrates a model square crystal lattice with a parallel alignment of spins, and the corresponding momentum-space electronic structure of a nodeless s-wave ferromagnetic phase. It comprises polarized majority-spin (spin-up) and minority-spin (spin-down) energy iso-surfaces which break the SO(3) spin-space symmetry, but have a shape which preserves the crystal-lattice symmetry (e.g. the four-fold crystal rotation symmetry in our model square lattice).    

A model crystal containing two such square lattices that are mutually shifted and have mutual  antiparallel alignment of their spins represents an antiferromagnet.  The spin-up and spin-down energy iso-surfaces preserve the four-fold crystal-rotation symmetry, as in the above ferromagnetic model. Unlike the ferromagnet, however,  the translation symmetry relating the opposite-spin lattices protects degeneracy of the opposite-spin energy iso-surfaces, and the electronic structure thus remains unpolarized as in the non-magnetic phase.

We now focus on symmetries generating the nodal magnetic ordering in altermagnets\cite{Smejkal2021a,Smejkal2022a}. A representative model, shown in Fig.~\ref{fig:AM-FM}a,  consists of two opposite-spin square lattices intertwined with a non-magnetic square lattice. The latter renders the sites hosting the two types of spins different (not related by translation), but related by a four-fold rotation.
The resulting spin density in this model altermagnet has a four-lobe d-wave form centered at the non-magnetic atom, and with the lobes having alternating signs  and separated by spin-degenerate nodes (see Fig.~\ref{fig:AM-FM}a and Fig.~\ref{fig:AM}b).   The same d-wave spin-density profile is repeated in the neighboring crystal unit cell. This ferroic nature of the ordering of the d-wave spin densities in the model altermagnet is reminiscent of the ferroic order of atomic magnetic dipoles in the ferromagnet (Fig.~\ref{fig:AM-FM}b).

The characteristic spin symmetry of the magnetic order  on the model crystal lattice in Fig.~\ref{fig:AM-FM}a combines a two-fold spin rotation $C_2$ around an axis orthogonal to the collinearity axis of the spins with the crystal four-fold rotation $C_4$. This spin symmetry, denoted as $[C_2||C_4]$, makes the altermagnet principally distinct from the ferromagnet which has no symmetry combining the spin-space rotation with any crystal symmetry transformation. 

In the momentum space, the spin-polarized d-wave band structure of the model altermagnet comprises opposite-spin energy iso-surfaces connected by the same $[C_2||C_4]$ symmetry, which enforces an equal number of spin-up and spin-down states, i.e., zero net magnetization. The magnetic order in the crystal space is also reflected by each spin-channel energy iso-surface in the momentum space  breaking the four-fold symmetry of the underlying crystal lattice (Fig.~\ref{fig:AM-FM}a). These even-parity distortions are mutually rotated between the opposite-spin iso-surfaces, resulting in spin-degenerate nodes and spin splitting away from the nodes with an even-parity alternation of the spin-splitting sign. In this regard, the nodal d-wave magnetic ordering in the altermagnet\cite{Smejkal2021a} is a realization of the nematic state both in the real space and in the spin space (the so called nematic-spin-nematic state) \cite{Kivelson2003,Wu2007}. 

For comparison, let us consider an alternative hypothetical scenario also respecting the $[C_2||C_4]$ symmetry. The spin splitting in this scenario would result in one larger and one smaller energy iso-surface, where each iso-surface keeps a shape preserving the crystal-lattice symmetry, in analogy to the nodeless ferromagnet in Fig.~\ref{fig:AM-FM}b. For this nodeless scenario, the altermagnetic  $[C_2||C_4]$ symmetry would have to be preserved separately in each of the two iso-surfaces, i.e., the sign of the spin polarization would have to alternate on each iso-surface.  However, for a collinear magnetic order protecting by symmetry a single momentum-independent spin quantization axis, i.e. excluding a spin texture with a momentum dependent magnitude or angle of the spin vector, this would lead to spin discontinuities on each iso-surface. The collinearity thus enforces the nodal form with distorted spin-up and spin-down iso-surfaces of the even-parity-wave ordering in altermagnets, as illustrated in Fig.~\ref{fig:AM-FM}a.

The above analysis can be cast in a rigorous systematic symmetry description of altermagnetism\cite{Smejkal2021a} using the framework of spin symmetry groups\cite{Smejkal2021a,Litvin1974,Litvin1977,Liu2021,McClarty2024,Xiao2023,Smolyanyuk2024,Shinohara2024,Watanabe2024,Jiang2023,Ren2023,Schiff2023}. Treating all collinear spin arrangements on crystals by the spin-group theory showed that their momentum and spin-dependent electronic structures fall into three distinct mutually exclusive classes of ferromagnets, antiferromagnets and altermagnets\cite{Smejkal2021a}. We now briefly summarize this classification.

A  collinear magnet remains invariant if the spins are rotated by any angle around the common spin-polarization axis, or if the spins are inverted (i.e. time-reversed) and then undergo a two-fold rotation around an axis orthogonal to the spins. For all collinear magnets, the so-called spin-only group contains these symmetries\cite{Smejkal2021a,Litvin1974,Litvin1977}, and we denote it as ${\bf r}_{\rm s}^{\rm cl}$. What distinguishes between ferromagnets, antiferromagnets and altermagnets are additional symmetries\cite{Smejkal2021a} forming  the  so-called non-trivial spin groups\cite{Smejkal2021a,Litvin1974,Litvin1977}. These symmetries contain pairs of, in general, different transformations applied simultaneously in the spin space and in the real space. We denote the non-trivial spin (Laue) groups ${\bf R}_{\rm s}^i$, where $i=$~I, II, III for the three distinct classes of ferromagnets, antiferromagnets and altermagnets, respectively\cite{Smejkal2021a}. Note that the Laue group is used instead of the point group because  ${\bf r}_{\rm s}^{\rm cl}$ ensures an even-parity band structure regardless of whether the magnetic crystal structure is inversion symmetric or not \cite{Smejkal2021a}. 

${\bf R}_s^{\rm I}=[E\parallel{\bf G}]$ corresponds to  the nodeless ferromagnets with split majority-spin and minority-spin bands. Here $E$ is a spin-space identity and {\bf G} is the crystallographic Laue group of real-space symmetry transformations. ${\bf R}_s^{\rm I}$ shows that the symmetries of  {\bf G} are preserved in each spin channel of the band structure.

In ${\bf R}_s^{\rm II}=[E\parallel{\bf G}]+[{C}_2\parallel{\bf G}]$, the first term again explicitly shows the preserved symmetries of  {\bf G} for each spin channel, while the second term, in combination with  ${\bf r}_{\rm s}^{\rm cl}$, protects spin degeneracy at all momenta, thus describing antiferromagnets. 

The only remaining symmetry class of collinear magnets, corresponding to the nodal altermagnets, is described by ${\bf R}_s^{\rm III}=[E\parallel{\bf H}]+[{C}_2\parallel A] \, [E\parallel{\bf H}]$. Here {\bf H} is a halving subgroup of the crystallographic Laue group {\bf G}, and $A$ can only be  real-space proper or improper rotation, but not real-space inversion\cite{Smejkal2021a}. Breaking of the symmetries of  {\bf G} in a given spin channel of the band structure is reflected explicitly by the first term in ${\bf R}_s^{\rm III}$. The second term protects zero net magnetization, while enabling the alternating spin polarization in the band structure. This, combined with the even parity, implies time-reversal symmetry breaking (TRSB) in the non-relativistic band structure of altermagnets in the absence of external magnetic field and internal magnetization\cite{Smejkal2021a,Reichlova2024,Betancourt2021}. 

There are 11 (32)  non-trivial spin Laue (point) groups in the ferromagnetic class, 11 (53) in the  antiferromagnetic class, and 10 (37) in the  altermagnetic class \cite{Smejkal2021a}. Among the altermagnetic spin Laue groups, 4 correspond  to d-wave ($l=2$) , 4 to g-wave ($l=4$) and 2 to i-wave ($l=6$) \cite{Smejkal2021a}. In the d, g, and i-wave altemagnets, bands are spin degenerate at 2, 4, and 6 nodal surfaces crossing the $\boldsymbol\Gamma$-point in the Brillouin zone. 

Altermagnetism is more abundant than ferromagnetism among the collinear magnets, and is frequently identified not only in metallic but also in insulating materials\cite{Smejkal2022a}.  Numerous altermagnetic candidate materials have been already identified based on their spin-group symmetries when analyzing, e.g., the published magnetic structures in the MAGNDATA database on the Bilbao Crystallographic Server \cite{Smejkal2021a, Smejkal2022a,Guo2023b,Xiao2023}. Besides 3D crystals, altermagnets can be also realized in 2D crystals\cite{Smejkal2022a,Smejkal2022GMR,Ma2021,Egorov2021,Brekke2023,Cui2023,Chen2023b,Mazin2023a,Sodequist2024} and, besides inorganics, also in organic materials\cite{Naka2019,Ferrari2024}.

\begin{figure}[h!]
	\centering
	\includegraphics[width=1\linewidth]{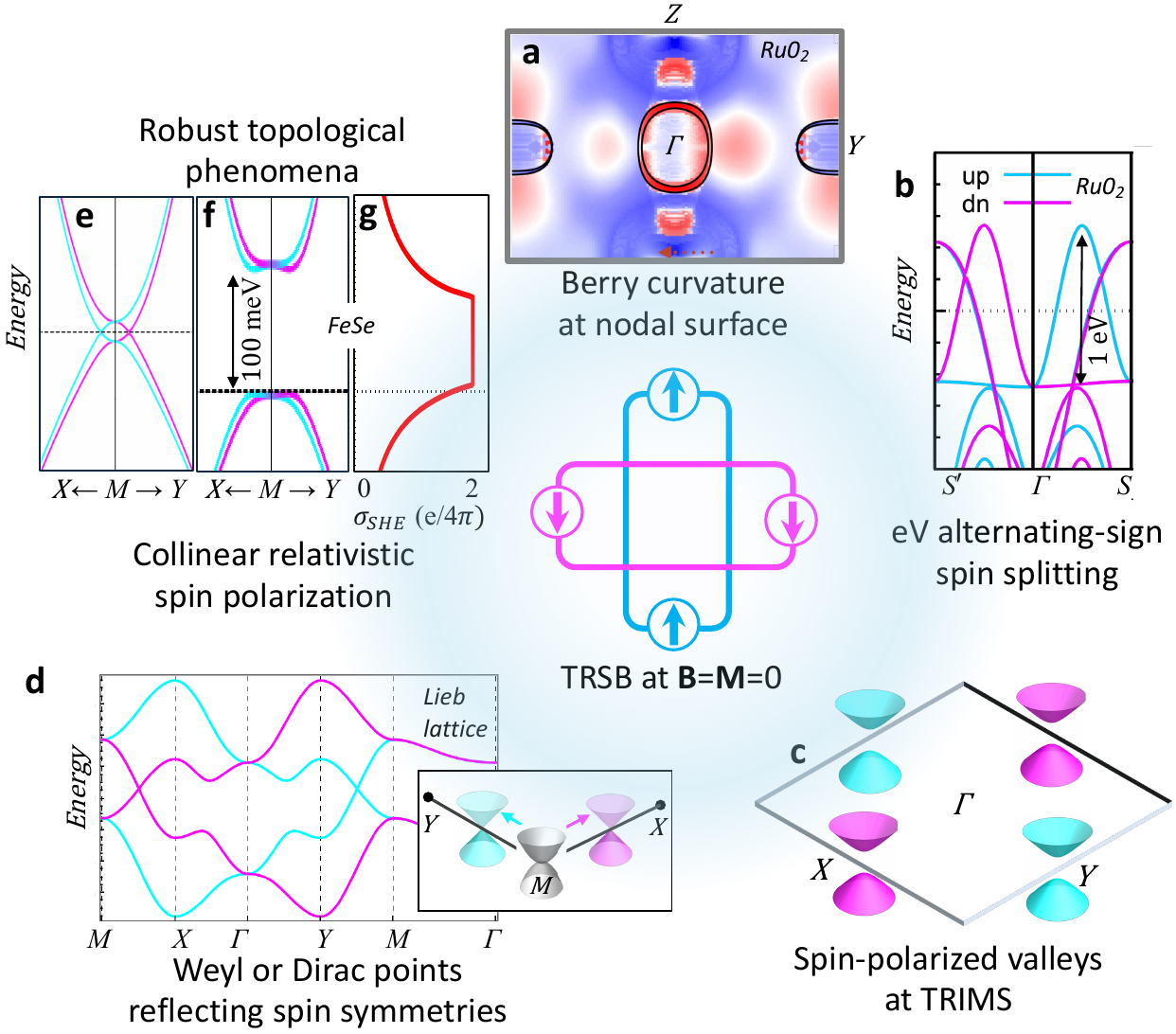}
	\caption{
	\textbf{Salient electronic structure features of altermagnets.}
Central panel highlight time-reversal symmetry breaking (TRSB) in the non-relativistic even-parity-wave electronic structure of altermagnets at zero external magnetic field and internal magnetization. {\bf a,} Berry curvature projected on the nodal surface with hot spots around the relativistic split bands intersecting the nodal surface. {\bf b,} Spin splitting with opposite sign along two perpendicular directions and magnitude reaching eV scale. DFT calculations are for the altermagnetic phase of RuO$_2$, and the panels {\bf a} and {\bf b} are adapted from Ref.~\onlinecite{Smejkal2020}. {\bf c,} Cartoon of non-relativistic spin-split valleys around time-reversal invariant momenta (TRIMS)\cite{Reichlova2024,Smejkal2022GMR,Ma2021,Cui2023,Zhu2024a}.  {\bf d,} Band structure of the altermagnetic 2D Lieb lattice model (Fig. 4b); the inset shows the emergence of symmetry-related Dirac points from the splitting of the quadratic band crossing. adapted from Ref.~\onlinecite{Antonenko2024}. {\bf e,}  DFT calculation of non-relativistic bands of 2D d-wave altermagnetic candidate FeSe. {\bf f,} Corresponding relativistic DFT calculation showing collinear  spin polarization in a topological insulating phase. {\bf g,} Corresponding quantum spin Hall effect with precise quantization of the spin-Hall conductivity. Panels {\bf e-g}  are adapted from Ref.~\onlinecite{Mazin2023a}.
}
\label{fig:Topo}
\end{figure}


Spin degeneracy at the nodal surfaces can be lifted by the relativistic spin-orbit coupling which, among a range of phenomena further reviewed in Sec.~D,  can generate Berry-curvature hot spots (Fig.~\ref{fig:Topo}a) and, correspondingly,  large values of the TRSB anomalous Hall effect. This was theoretically predicted in  altermagnetic candidates RuO$_2$  \cite{Smejkal2020} or FeSb$_2$\cite{Mazin2021}, and reviewed in Ref.~\onlinecite{Smejkal2022AHEReview}. Besides RuO$_2$, initial experimental studies reported the anomalous Hall effect also in altermagnetic candidates MnTe and Mn$_5$Si$_3$\cite{Feng2022,Tschirner2023,Wang2023a,Betancourt2021,Reichlova2024,Han2024,Kluczyk2023}. 

Optical and X-ray magnetic circular dichroisms\cite{Samanta2020,Zhou2021a,Hariki2023}, which are the ac counterparts of the dc anomalous Hall effect, were employed to detect N\'eel vector reversal in MnTe and Mn$_5$Si$_3$ \cite{Hariki2023,Han2024}. X-ray magnetic circular dichroism was also employed in combination with photoemission electron microscopy to perform a high-resolution vector mapping  of real-space altermagnetic configurations  in MnTe, ranging from nano-scale vortices  and domain walls to micron-scale single-domain states \cite{Amin2024}. A thermoelectric counterpart of the anomalous Hall effect -- the anomalous Nernst effect\cite{Zhou2023a}, was reported in experiments in the altermagnetic candidate Mn$_5$Si$_3$\cite{Badura2024,Han2024a}.

The possibility of strong spin polarization in the non-relativistic electronic structure  away from the nodal surfaces, with spin splitting of the bands on an eV scale,  was pointed out in the initial theoretical studies of  RuO$_2$\cite{Smejkal2020,Ahn2019} (Fig.~\ref{fig:Topo}b). The presence of spin-polarized valleys  (Fig.~\ref{fig:Topo}c), which in altermagnets can form at time-reversal invariant momenta (TRIM), is another distinctive band-structure feature highlighted in theoretical studies of several altermagnetic candidates\cite{Reichlova2024,Smejkal2022GMR,Ma2021,Cui2023,Zhu2024a}.  

The spin splitting and TRSB in non-relativistic metallic band structures was proposed to generate altermagnetic counterparts of the giant magnetoresistance \cite{Smejkal2022GMR} and tunneling magnetoresistance \cite{Smejkal2022GMR,Shao2021,Jiang2023a,Shao2023,Cui2023,Samanta2024,Chi2024} effects, and of the longitudinal and transverse spin-current and spin-transfer torque effects\cite{Gonzalez-Hernandez2021,Naka2021,Bose2022,Bai2022,Karube2022,Shao2023,Cui2023}, as reviewed in Ref.~\onlinecite{Smejkal2022a}. In ferromagnets, these phenomena underpin the functionality of spintronic memory technologies which are now starting to replace semiconductor non-volatile memories on advanced-node processor chips. While the superior scalability of the spintronic memory bits compared to the semiconductor embedded-memory technologies has enabled the present breakthrough,  future spatial, temporal and energy scalability of  spintronic memories will be limited by the magnetization of the employed ferromagnetic components. The strong spin-polarization phenomena accompanied by the absence of net magnetization in altermagnets are thus a major incentive for investigating their possible use in future spintronic technologies. 



\begin{figure}[h!]
	\centering
	\includegraphics[width=.8\linewidth]{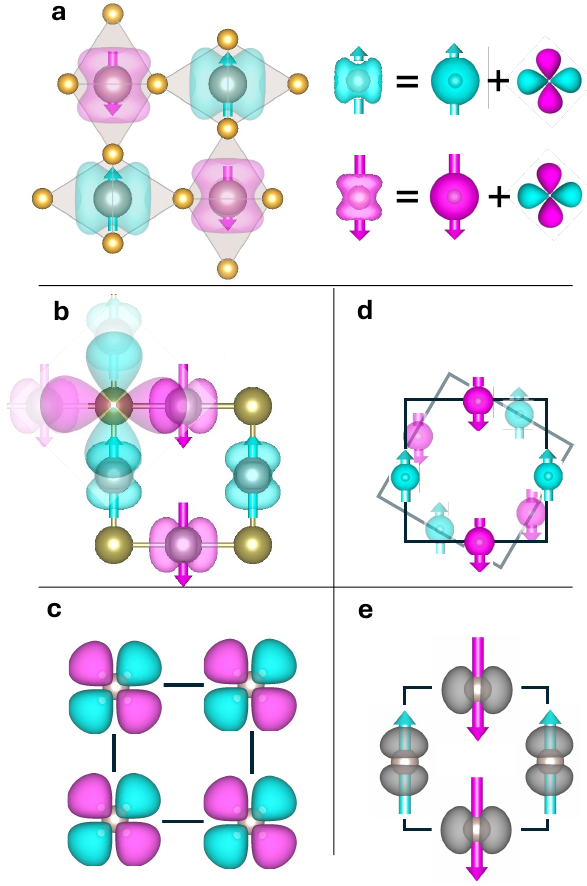}
	\caption{
	\textbf{Realizations of altermagnetic order in crystal lattices.}
{\bf a,} Rutile lattice \cite{Smejkal2020} with spin densities on magnetic atoms decomposed into s-wave (dipole) and d-wave components. {\bf b,} Lieb lattice  \cite{Brekke2023,Antonenko2024} showing an effective d-wave spin density centered at the non-magnetic atom. {\bf c,} d-wave spin density on a magnetic atom featuring no atomic dipole component. {\bf d,e} Altermagnetic symmetries generated by lattice distortions (e.g. twists)\cite{Mazin2023a,Chakraborty2024,Liu2024} or orbital ordering \cite{Leeb2023}.
}
\label{fig:AM}
\end{figure}

\begin{figure}[h!]
	\centering
	\includegraphics[width=1\linewidth]{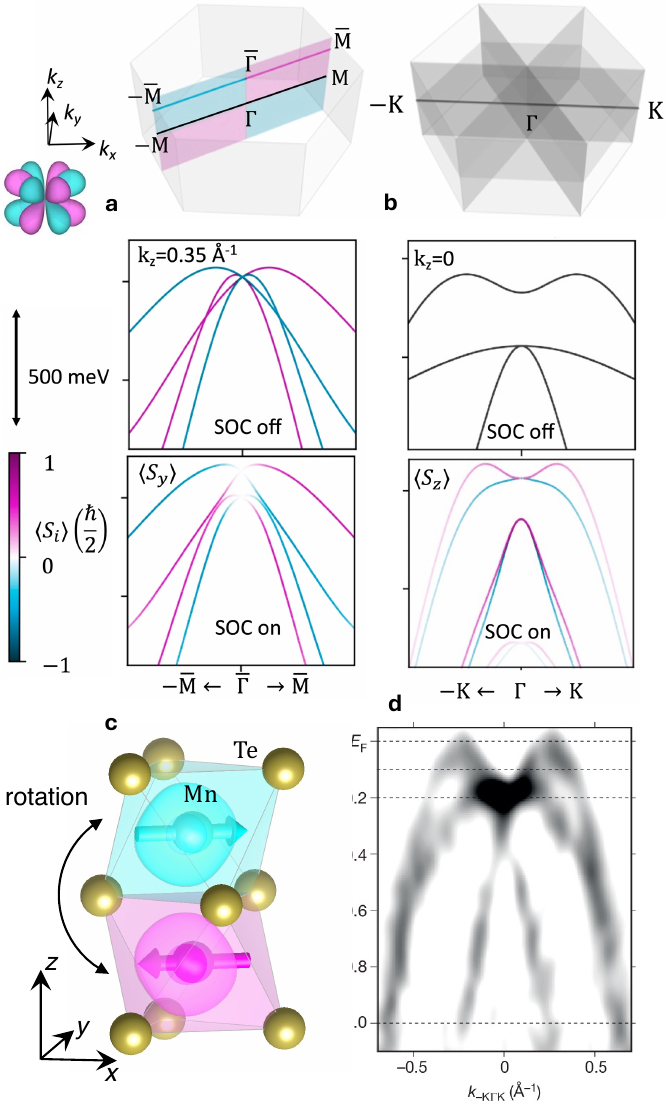}
	\caption{
	\textbf{Altermagnetic MnTe.}
{\bf a,} DFT spin-polarized band structure along a path in the Brillouin zone, depicted in the upper inset,  away from the four nodal planes of the g-wave altermagnet MnTe.  The top and bottom DFT panels are with relativistic spin-orbit coupling (SOC) turned off and on, respectively. {\bf b,} Same as {\bf a} but along a path in the $k_z=0$ nodal plane.  {\bf c,}  MnTe crystal structure with the opposite-spin Mn sublattices connected by crystal rotation. {\bf d,} Angle-resolved photoemission measurements of the band structure corresponding to the bottom DFT panel in {\bf b}. Adapted from Ref.~\onlinecite{Krempasky2024}.
}
\label{fig:MnTe}
\end{figure}

\begin{figure}[h!]
	\centering
	\includegraphics[width=1\linewidth]{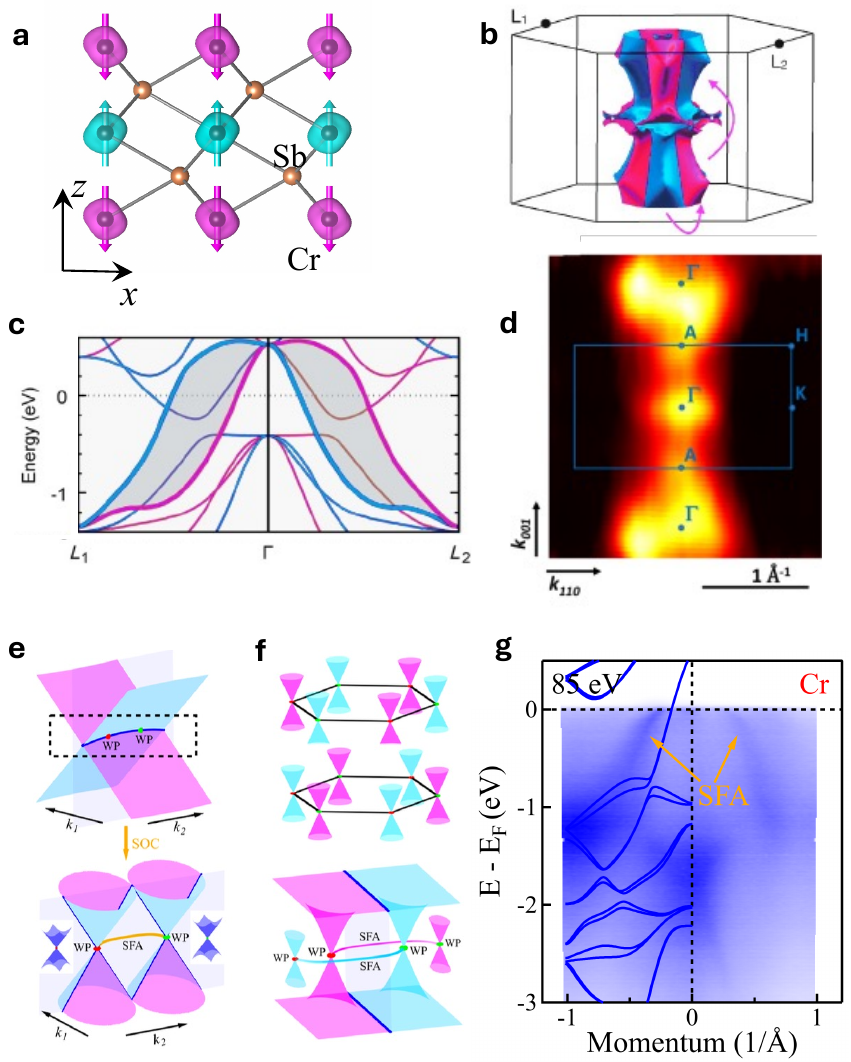}
	\caption{
	\textbf{Altermagnetic CrSb.}
{\bf a,} Crystal structure with DFT spin densities around the Cr atoms. {\bf b,} Non-relativistic DFT Fermi surface with four nodal planes corresponding to the g-wave ordering. {\bf c,} Non-relativistic DFT spin-split bands away from the nodal planes. Panels {\bf a}-{\bf c} are adapted from Ref.~\onlinecite{Smejkal2021a}. {\bf d,} Angle-resolved photoemission measurements of the Fermi surface corresponding to the DFT panel {\bf b}. Adapted from Ref.~\onlinecite{Reimers2024}. {\bf e,f,} Cartoons of Weyl points and surface Fermi arcs (SFA) corresponding to crossings of opposite-spin and same-spin bands, respectively.  The Weyl points are not located along high-symmetry directions. {\bf g,} Angle-resolved photoemission measurement of SFA. Panels {\bf e}-{\bf g} are adapted from Ref.~\onlinecite{Li2024}. 
}
\label{fig:CrSb}
\end{figure}

We conclude this section with Fig.~\ref{fig:AM} where we show, using cartoons of spin densities in the direct crystal space, a range of demonstrated or predicted physical realizations of altermagnetism. Fig.~\ref{fig:AM}a illustrates on the rutile crystal structure cases where the ${\bf R}_s^{\rm III}$ symmetry is realized by the arrangement of local anisotropic atomic spin densities.  These spin densities can be decomposed into s-wave (dipole) components with antiferroic alignment between neighboring atoms, and d-wave (or higher even-parity-wave)  components aligned ferroicly in the crystal lattice. The higher even-parity-wave spin density components correspond to the higher even-parity-wave symmetry of anisotropic exchange interactions in the crystal lattice\cite{Smejkal2023}. This scenario applies to many identified material candidates  for the  altermagnetic phase\cite{Smejkal2022a}. As examples of altermagnets whose corresponding spin-polarized electronic structures were confirmed by spectroscopic measurements \cite{Krempasky2024,Lee2024,Osumi2024,Hajlaoui2024,Reimers2024,Yang2024,Ding2024,Li2024,Lu2024},  we show in Figs.~\ref{fig:MnTe} and \ref{fig:CrSb}  the room-temperature altermagnetic semiconductor MnTe \cite{Smejkal2021a,Betancourt2021,Lovesey2023b,Mazin2023,Krempasky2024,Lee2024,Osumi2024,Hajlaoui2024,Grzybowski2024,Aoyama2024,Kluczyk2023}, and  the metallic altermagnet CrSb  with the ordering temperature above 700~K and spin-splitting magnitudes exceeding 1~eV \cite{Smejkal2021a,Reimers2024,Yang2024,Ding2024,Li2024,Lu2024}.  We note that experimental signatures of magnetism in RuO$_2$, a metallic rutile crystal considered among this class of altermagnets from the early theoretical studies  \cite{Smejkal2020}, are a matter of an ongoing debate \cite{Berlijn2017a,Zhu2018,Lovesey2022,Occhialini2021,Feng2022,Bose2022,Bai2022,Karube2022,Lovesey2023c,Liu2023,Fedchenko2024,Smolyanyuk2024a,Lin2024,Kessler2024,Li2024a,Wenzel2024,Jeong2024}.  In other insulating rutile crystals\cite{Smejkal2022a,Noda2016,Hayami2019,Yuan2020,Bhowal2024}, experimental signatures of magnetism are well established.

Another realization of altermagnetism is illustrated in Fig.~\ref{fig:AM}b on a Lieb lattice with a compensated collinear magnetic order\cite{Brekke2023,Antonenko2024}. Here the spin density of neighboring magnetic atoms forms directly an effective d-wave (or higher even-parity-wave) pattern with the origin at the central non-magnetic atom.
Apart from this scenario, it is anticipated that the ${\bf R}_s^{\rm III}$ symmetries can be also realized by  ferroic ordering of local d-wave (or higher even-parity-wave) spin densities on magnetic atoms without any local atomic dipole components, as illustrated in Fig.~\ref{fig:AM}c. 

Fig.~\ref{fig:AM}d shows an illustration of predicted realizations of altermagnetic symmetries by crystal lattice deformations\cite{Mazin2023a,Chakraborty2024,Liu2024}. While in the unperturbed crystal the opposite spins are related by one of the non-altermagnetic symmetries (translation or inversion), the deformation (e.g. a twist of crystal planes) breaks these symmetries while preserving the characteristic spin and crystal rotation symmetry $[{C}_2\parallel A]$ of altermagnets. 

Finally, Fig.~\ref{fig:AM}e illustrates a case where the role of symmetries of single-particle potentials of the ionic crystal lattice are complemented by electronic correlations in the formation of the altermagnetic phase. The spin arrangement on the ionic crystal lattice is of type ${\bf R}_s^{\rm II}$ which prohibits altermagnetism. The altermagnetic phase is then enabled by correlation-induced orbital ordering  \cite{Leeb2023,Fernandes2024} which changes the symmetry to ${\bf R}_s^{\rm III}$. 


\subsection{Relativistic spin-orbit coupling, topological, and correlation phenomena in altermagnetic band-structures}
\label{relativistic} 


In this section we focus on extraordinary relativistic and topological phenomena in the electronic structure of altermagnets\cite{Krempasky2024,Fakhredine2023,Mazin2023a,Fang2023a,Li2024b,Fernandes2024,Antonenko2024,Nag2023,Roig2024,Parshukov2024,Rao2024,Li2024,Lu2024,Zhu2023d,Ghorashi2023,Zhao2024}. In the last part of the section, we add comments on correlation effects \cite{Leeb2023,Fernandes2024,Roig2024,Das2023,Maier2023,Sato2023,Bose2024,Ferrari2024,Rooj2024,Bernardini2024}. 

We start with the relativistic spin-orbit coupling. Unlike the spontaneous symmetry-breaking by the altermagnetic ordering in the  ground state, discussed in the previous section, the relativistic spin-orbit coupling is a symmetry-breaking single-particle term present already in the Hamiltonian of the system. Because it arises  in the $1/c^2$ expansion beyond the non-relativistic limit of the Dirac equation, where $c$ is the speed of light, it is  a typically weak perturbative correction to the electron-electron interaction and ionic-potential terms in the Hamiltonian. Despite this, it generates a plethora of remarkable phenomena in non-magnetic and magnetic systems utilized, e.g., in spintronics\cite{Manchon2019}. This phenomenolgy is further enriched by extraordinary interplays of the spin-orbit coupling with the altermagnetic ordering.


In the following paragraphs, we give an example of such an extraordinary interplay on relativistic bands of MnTe and FeSe 3D and 2D altermagnets\cite{Krempasky2024,Mazin2023a}. The feature that we highlight  is a possibility to realize  in altermagnets relativistic band structures in which the spin-polarization axis is momentum-independent across high-symmetry directions, planes or the entire Brillouin zone. This is extraordinary given the general form, $\sim {\bf s}\cdot({\bf k}\times{\bf E})$, of the spin-orbit coupling term arising from the Dirac equation, where {\bf s} denotes spin, {\bf k} momentum and {\bf E} electric field. The coupling between spin and momentum vectors implies that relativistic systems tend to feature spin textures in their electronic structure where the magnitude and direction of spin varies  with momentum, as commonly observed in ferromagnets or non-centrosymmetric non-magnetic materials. 

We now contrast the relativistic spin textures with the effect of the spin-orbit coupling on  the  $k_z=0$ nodal plane in altermagnetic MnTe. This material has a non-relativistic spin-polarized band structure of the nodal g-wave type, thus displaying 4 nodal planes crossing the $\boldsymbol\Gamma$-point (see Fig.~\ref{fig:MnTe}b). The spin degeneracy, protected at the nodal planes by the non-relativistic spin symmetry, can be lifted by the relativistic spin-orbit coupling \cite{Krempasky2024}. Remarkably, the spin-orbit coupling can spin-split the bands despite the inversion symmetry of MnTe. This already indicates the distinct phenomenology of spin-orbit coupling effects in altermagnets, compared to conventional spin splitting by the spin-orbit coupling which requires broken inversion symmetry. 

When the N\'eel vector in MnTe is in the magnetic easy-plane ($c$-plane of the MnTe crystal), the strong non-relativistic g-wave order generates a corresponding in-plane spin-polarization component along the N\'eel vector away from the nodal planes. This is complemented on the $k_z=0$ nodal plane by a relativistic out-of-plane spin-polarization whose sign alternates but the polarization axis is independent of the in-plane momentum (Fig.~\ref{fig:MnTe}b) \cite{Krempasky2024}. We point out that the extraordinary absence of a momentum-dependent spin texture on the $k_z=0$ plane is realized at energies in the band structure with a strong admixture of orbitals from the heavy element Te. Indeed the relativistic spin splitting in this part of the band structure reaches large magnitudes $\sim 100$~meV (Fig.~\ref{fig:MnTe}b,d). 

The realization of such a common momentum-independent spin-polarization axis has been a long-sought goal in the research of relativistic band-structure effects. Before altermagnets, this was only observed in non-magnetic 2D semiconductors with fine-tuned strengths of microscopic Rashba and Dresselhaus spin-orbit coupling \cite{Bernevig2006,Koralek2009}. In altermagnetic MnTe,  the collinear relativistic spin-polarization on the $k_z=0$ nodal plane, with all spins pointing along the  normal to the plane, is symmetry protected\cite{Krempasky2024}. Specifically, the absence of any in-plane spin-polarization component on the $k_z=0$ nodal plane is enforced by the relativistic non-symmorphic mirror symmetry where the mirror plane is parallel to the crystal c-plane. 

We emphasize that this relativistic symmetry is only present for the N\'eel vector oriented along the c-plane in the crystal\cite{Krempasky2024}. In contrast, in the g-wave altermagnet CrSb, which has identical crystallographic and non-relativistic spin groups as MnTe\cite{Smejkal2021a}, the relativistic spin-orbit coupling leads to the N\'eel vector easy-axis pointing along the crystal c-axis. This changes the relativistic symmetry from the non-symmorphic c-plane mirror in MnTe to a combined non-symmorphic c-plane mirror with time-reversal in CrSb. This CrSb symmetry does not enforce the absence of the in-plane  spin-polarization component in the $k_z=0$ plane; instead it enforces the absence of the out-of-plane spin-polarization component along the $k_x=k_y=0$ line (which turns out to be a completely spin-degenerate nodal line because of the interplay with another relativistic mirror symmetry)\cite{Fernandes2024}.

An additional extraordinary feature of the relativistic spin-orbit coupling in the MnTe altermagnet with in-plane N\'eel vector is a quadratic band dispersion and spin splitting around the $\boldsymbol\Gamma$-point (Fig.~\ref{fig:MnTe}b,d). The absence of the constant and linear spin-splitting terms highlights the principal distinction  from the exchange spin splitting present in ferromagnets and the relativistic spin splitting present in the non-magnetic non-centrosymmetric crystals. 

The collinear relativistic spin-polarization is not a feature seen exclusively in the band structure of MnTe. In Fig.~\ref{fig:Topo}e,f we show the electronic structure of a candidate 2D altermagnet FeSe \cite{Mazin2023a}. In the non-relativistic limit (Fig.~\ref{fig:Topo}e), the spin-polarized 2D bands have a d-wave ordering. When spin-orbit coupling is included and the N\'eel vector is oriented in the direction normal to the 2D plane (Fig.~\ref{fig:Topo}f), spin-polarized states in valleys around {\bf M}-points in the Brillouin zone acquire a common spin axis, this time parallel to the N\'eel vector. For energy ranges where the spectrum contains only the {\bf M}-point valleys, the entire energy iso-surface in the 2D momentum space has a common momentum-independent spin-polarization axis. This is again despite the large spin-orbit-coupling strength introduced by Se, which generates a splitting in the {\bf M}-point valleys  on the $\sim 100$~meV scale (Fig.~\ref{fig:Topo}f).


We now move on to the topological phenomena in the altermagnetic band structures which are also unparalleled in ferromagnetic or non-magnetic systems \cite{Mazin2023a,Ghorashi2023,Fang2023a,Li2024b,Fernandes2024,Antonenko2024,Nag2023,Roig2024,Parshukov2024,Rao2024,Li2024,Lu2024,Li2023,Zhu2023d,Ghorashi2023,Zhao2024}. Several of these phenomena are related to the spin-degenerate nodal lines in the Brillouin zone, which are the remnants of the non-relativistic nodal planes of the altermagnet when the relativistic spin-orbit coupling is included. They are topologically trivial with respect to non-spatial symmetries. When present, however, these Brillouin-zone nodal lines, and the corresponding Weyl nodes in the band structure, can be protected by mirror symmetries of the crystal. Therefore, they remain stable against small perturbations (e.g. by magnetic field or strain) that preserve the mirror symmetries \cite{Fernandes2024}. 

We again illustrate these and other topological phenomena  on representative altermagnetic materials. In CrSb, the splitting of the non-relativistic $k_z = 0$ nodal plane by the relativistic spin-orbit coupling leads to the emergence of pairs of Weyl points and thus Fermi surface arcs (Fig.~\ref{fig:CrSb}e)\cite{Li2024}. Here the Weyl points result from the crossing of  bands with opposite spin. Conversely, along the spin-split parts of the Brillouin zone in the non-relativistic band structure, crossings of  bands with the same spin give rise to spin-polarized Weyl points (Fig.~\ref{fig:CrSb}f). Fermi arcs connecting the surface projections of the Weyl points with the same total spin are also spin polarized (Fig.~\ref{fig:CrSb}f)\cite{Li2024,Lu2024}.

An extraordinary interplay of topology and relativistic spin-orbit coupling can be illustrated on the 2D altermagnetic candidate FeSe (Fig.~\ref{fig:Topo}e-g)  \cite{Mazin2023a}. The non-relativistic bands feature spin-degeneracy above and below the Fermi level at the {\bf M}-points which are located at the two nodal lines in the 2D Brillouin zone of this d-wave altermagnet (Fig.~\ref{fig:Topo}e). In addition, there are two crossings of bands with the same spin  at the Fermi level. They are located on {\bf M}-{\bf X} and {\bf M}-{\bf Y} lines (Fig.~\ref{fig:Topo}e), respectively, and are connected by the alternagnetic spin symmetries.  Spin-orbit coupling splits these same-spin band crossings and shifts further apart in energy the two spin-degenerate bands at the {\bf M}-point by $\sim 100$~meV, resulting in the formation of a 2D topological spin-Chern insulator  (Fig.~\ref{fig:Topo}f). As already highlighted above in the discussion of the  spin-orbit coupling effects in FeSe,  the spin polarization maintains a momentum-independent axis in the valleys around the {\bf M}-points, even for this relatively large strength of the spin-orbit coupling in FeSe.  This leads to an exceptionally precise quantization of the quantum spin Hall effect over a broad range of energies (Fig.~\ref{fig:Topo}g)\cite{Mazin2023a}.

Crossings of bands with the same spin, and their connection to non-trivial topology, are not particular to monolayer FeSe or CrSb, but are widespread in altermagnetic materials \cite{Antonenko2024,Parshukov2024,Roig2024}. Further insights about the topological nature of such crossings follow from analyzing the 2D Lieb lattice model (Fig.~\ref{fig:AM}b) \cite{Antonenko2024}, of which monolayer FeSe is a particular material realization. A characteristic feature of this model, illustrated in Fig.~\ref{fig:Topo}d, is the existence of a quadratic band crossing  at the {\bf M}-point \cite{Sun2009}. Altermagnetic order splits the quadratic band crossing into pairs of crossings of same-spin bands (Dirac points in 2D) located at the {\bf M}-{\bf X} and {\bf M}-{\bf Y} zone edges and related by the $[C_4||C_2]$ spin symmetry (see inset of Fig.~\ref{fig:Topo}d). While these Dirac points are guaranteed to exist for an infinitesimally small sublattice magnetic moment, a large enough moment will bring the Dirac crossings to the {\bf X} and {\bf Y} points and remove them. The impact of the spin-orbit coupling depends on the direction of the N\'eel vector\cite{Antonenko2024}. For the out-of-plane direction, like in the case of the above discussed FeSe, the Dirac points are gapped out, resulting in the topological quantum spin Hall state and mirror spin-Chern bands.\cite{Mazin2023a,Antonenko2024}. For the in-plane N\'eel vector, however, even if the Dirac points are gapped, the bands are topologically trivial. This illustrates that the direction of the N\'eel vector has a substantial impact on the opposite-spin and same-spin band crossings. In 3D lattices, same-spin band crossings can also occur, where they can give rise to Weyl nodal loops in the presence of spin-orbit coupling \cite{Antonenko2024}.

Another promising route to imprint unusual electronic properties in altermagnetic materials is to combine their characteristic nodal magnetic ordering with other phases, such as quantum anomalous Hall Chern insulators, axion insulators, multiferroics  or superconductors\cite{Smejkal2022a,Beenakker2023,Brekke2023,Li2023,Zhu2023d,Sumita2023,Ghorashi2023,Papaj2023a,Wei2024,Zhang2024a,Cheng2024,Zhao2024,Maeland2024,Chakraborty2024b,Banerjee2024a,Jeschke2024,Verbeek2023,Guo2023c,Bernardini2024,Zyuzin2024,Sim2024,Chakraborty2024a}. 
The interplay between superconductivity and altermagnetism has been investigated in the context of heterostructures, where pairing is induced by the proximity effect, and bulk systems, where pairing and altermagnetic ordering coexist. These studies have revealed the emergence of intriguing phenomena, such as unconventional Andreev reflection, pair density waves (first discussed in the context of the spin-polarized $l=2$ Pomeranchuk instability \cite{Soto-Garrido2014}), non-trivial topological modes, and non-reciprocal supercurrents.

So far we have mostly focused on the characteristics of the electroníc band structures of altermagnets. However, once electron-electron interactions become sufficiently strong, an effective single-particle band-structure description is no longer sufficient to capture the collective electronic properties. Besides substantial band-mass renormalizations, the quasiparticle lifetime changes sharply with energy and momentum. The role of electronic correlations in promoting or affecting altermagnetism has been studied in various contexts \cite{Das2023,Maier2023,Leeb2023,Sato2023a,Roig2024}, including in the strongly-coupled regime where the interaction is comparable to the bandwidth \cite{Bose2024,Ferrari2024}. 

Since the strongly-correlated Mott-insulator crystals commonly display a compensated antiparallel magnetic order, several Mott insulating materials with appropriate spin symmteries have been put forward as altermagnetic candidates. Specifically, a wide class of Mott insulating perovskites, such as the manganite CaMnO$_3$ and the titanate LaTiO$_3$ \cite{Fernandes2024,Bernardini2024,Rooj2024}, have been identified as platforms to realize and investigate altermagnetism in the regime of strong correlations. In these compounds, the oxygen octahedra rotate to accommodate the cation, lowering the ideal cubic symmetry of the perovskite down to orthorhombic, which essentially changes the spin group from class ${\bf R}_s^{\rm II}$ to class ${\bf R}_s^{\rm III}$. The octahedra rotation directly impacts orbital degrees of freedom, which in turn affect the magnetic interactions. Similarly, Mott insulating states realized in oxides with other Ruddlesden-Popper phases, such as the nickelate La$_3$Ni$_2$O$_7$ \cite{Bernardini2024} and the cuprate La$_2$CuO$_4$ \cite{Smejkal2021a}, possess the spin-symmetry requirements to display altermagnetism. Besides 3D oxides, 2D organic charge-transfer salts such as $\alpha$- and $\kappa$-(BEDT-TTF)$_2$X can also realize a Mott insulating altermagnetic phase due to the fact that their organic molecules can form an arrangement similar to the Shastry-Sutherland lattice\cite{Ferrari2024}.  
  

\subsection{Non-collinear spin arrangements on crystals and p-wave magnetism}
\label{odd-parity}
So far our focus was on  the collinear altermagets.  In this section we broaden our analysis of compensated magnetic phases by including non-collinear spin arrangements on crystals. Using the spin symmetries, we will show how salient features of the non-relativistic electronic structure of different types of non-collinear magnets depart from those of the  collinear altermagnets. We first give examples of non-collinear magnets with even-parity spin-split bands that break time-reversal symmetry. In the second part, we discuss non-collinear spin arrangements on crystals generating odd-parity time-reversal-symmetric band structures.

In general, non-relativistic band structures of non-collinear magnets have momentum-dependent magnitude or direction of the expectation value of spin. Moreover, also unlike the collinear magnets, the bands are inversion symmetric only for centrosymmetric magnetic crystal structures. Magnetically-ordered crystals from the Mn$_3$X family \cite{Chen2014,Kubler2014,Nakatsuji2015} are examples of non-collinear coplanar centrosymmetric magnets. They break non-relativistic spin symmetries combining the two-fold spin-space rotation  with real-space translation or inversion, and have a symmetry combining spin-space and real-space rotations (specifically the $[C_3||C_6]$ symmetry). As a result, their even-parity non-relativistic bands are spin split, while having zero net magnetization, in analogy to altermagnets. However, in contrast to altermagnets, the Fermi surfaces of the non-collinear  Mn$_3$X magnets feature a non-collinear spin texture. Also in contrast to the nodal altermagnets, and reminiscent of the nodeless ferromagnets, the shape of the Fermi surface preserves the symmetry of the underlying crystal lattice. Unlike the d-wave altermagnets, the non-collinear Mn$_3$X magnets are thus not a realization of the nodal anisotropic nematic-spin-nematic state\cite{Kivelson2003,Wu2007}.

We now move on to the odd-parity spin-split band structures. First, we highlight the case of non-centrosymmetric non-collinear coplanar magnets with a symmetry $[C_2||{\bf t}]$, combining a two-fold spin-space rotation around the axis orthogonal to the coplanar spins with a real-space translation. The upper right panel of Fig.~\ref{fig:SC-M} illustrates an example of magnetic structures that have this symmetry\cite{Hellenes2023}. They allow for a magnetic phase with a zero net magnetizetion and odd-parity spin-polarized  energy bands, $E_{\sigma}({\bf k})=E_{-\sigma}(-{\bf k})$. Such a p-wave magnetic phase, featuring parity-breaking spin-polarized Fermi surfaces that shift in opposite directions in the momentum space for opposite spin directions (top-right panel of Fig.~\ref{fig:SC-M}), has been predicted to be realized in the non-centrosymmetric non-collinear coplanar magnet CeNiAsO\cite{Hellenes2023}. 

The  direct-space $[C_2||{\bf t}]$ symmetry implies that the only allowed spin-polarization component in the momentum space for all momenta in the above p-wave magnets is along the $C_2$ rotation axis\cite{Hellenes2023}. Remarkably, this unique spin-polarization axis in the momentum space is perpendicular to the plane of the coplanar spins in the direct crystal space. This is to be contrasted with the case of collinear ferromagnets and altermagnets, whose momentum-independent spin-quantization axis in the band structure is oriented along the axis of the collinear spins in the crystal lattice. 

All non-collinear coplanar magnetic crystals remain invariant if the spins are inverted (i.e. time-reversed) and then undergo the  $C_2$ rotation around the axis orthogonal to the coplanar spins. This coplanarity symmetry,  denoted as $\bar{C}_2$, combined with the $[C_2||{\bf t}]$ symmetry, implies a symmetry combining time-reversal with the translation {\bf t}. The momentum-space band structure of the above p-wave magnets is, therefore, time-reversal symmetric.

Finally, we point out that non-collinear non-coplanar magnetic crystals with the symmetry combining time-reversal with translation can also have time-reversal symmetric spin-split bands if the magnetic crystal is non-centrosymmetric. However, because of the absence of the coplanarity symmetry $\bar{C}_2$, the bands have spin textures with, in general, momentum-dependent magnitude and direction of spins\cite{Hellenes2023}. These odd-parity time-reversal invariant spin-textures in the momentum space represent a non-relativistic magnetic counterpart of the relativistic spin-orbit coupled textures in non-centrosymmetric crystals in the normal non-magnetic phase.

\subsection{Summary}
\label{summary}
We conclude by summarizing the key points covered in this review. Altermagnets are characterized by a nodal even-parity-wave magnetic ordering in both the real-space crystal structure and the momentum-space electronic structure. The d-wave altermagnets can be regarded as realizations of the nematic state both in the real space and in the spin space, and as magnetic counterparts of unconventional d-wave superconductors. 

Altermagnetism arises from  the interplay of single-particle potentials of the crystal lattice and electron-electron interactions, and can be identified in numerous insulating and metallic materials based solely on symmetries of the spin densities in the crystal lattice. This is a distinct mechanism from a purely electronic instability of the correlated Fermi liquid (spin-polarized even-parity-wave Pomeranchuk instability), although the resulting symmetries of the magnetic ordering in the momentum space are analogous.

The non-relativistic electronic structure of the collinear altermagnets comprises separate spin-up and spin-down channels, with distorted energy iso-surfaces in each spin channel  breaking the symmetry of the underlying crystal lattice. The even-parity distortions are mutually rotated between the opposite-spin iso-surfaces, resulting in spin-degenerate nodes, and in alternating-sign even-parity spin splitting away from the nodes. The  spin-up and spin-down bands are degenerate at 2, 4 or 6 nodal surfaces in the Brillouin zone, protected by non-relativistic spin symmetries, and corresponding to d, g, or i-wave magnetic ordering. The even-parity-wave ordering breaks  time-reversal symmetry in the electronic structure in the absence of external magnetic field and internal magnetization. 

The salient band structure features of altermagnets can have unparalleled practical merits, e.g., in spintronics. They also underpin extraordinary relativistic and topological phenomena, such as collinear relativistic spin polarization or topological Weyl points that reflect the spin symmetries. In addition, altermagnets represent a unique platform to search for quantized topological responses at zero magnetic field and high temperatures, or to  combine this unique type of magnetic ordering with superconductivity.

Materials hosting odd-parity-wave magnetic ordering are predicted by employing an analogous framework of spin symmetries for the case of non-collinear magnetic configurations. They are expected to enrich the above altermagnetic research directions by the distinct phenomenology of their time-reversal symmetric  spin-polarized bands. The study of odd-parity-wave magnets may also bring new perspective on long-term open questions in quantum mechanics regarding the possible existence of ordered many-body ground states with equilibrium charge or spin currents. 

Notwithstanding the remarkable and rapid progress in the field, several open questions remain and new research directions emerge on the horizon. For instance, it is an intriguing question whether an altermagnetic quantum critical point can promote new phenomena\cite{Steward2023}, perhaps related to non-Fermi liquid behavior and superconductivity, that is not encountered in other widely studied quantum-critical ferroic orders, like ferromagnetism, ferroelectricity, and nematicity. The impact of correlations and bad metal behavior on altermagnetic properties of strongly-correlated materials is so far also a virtually unexplored direction. Finally, an extensive experimental effort is needed to establish all the predicted extraordinary electronic structure features of altermagnets, to explore the corresponding unconventional electronic and optical responses, and to exploit them in novel devices.







\end{document}